\documentclass{article}
\usepackage{arxiv}
\usepackage[utf8]{inputenc} 
\usepackage[T1]{fontenc}    
\usepackage{hyperref}       
\usepackage{url}            
\usepackage{booktabs}       
\usepackage{amsfonts}       
\usepackage{nicefrac}       
\usepackage{microtype}      
\usepackage{lipsum}		
\usepackage{graphicx}
\usepackage{doi}

\usepackage{hyperref}

\usepackage{biblatex}
\hypersetup{
    colorlinks=true,
    linkcolor=blue,
    filecolor=magenta,      
    urlcolor=cyan,
    pdftitle={Overleaf Example},
    pdfpagemode=FullScreen,
}

\author{ Colin Topping \\
	Bristol Cyber Security group\\
	University of Bristol, UK\\
	\\	
    \And
	Ola Michalec \\
	Bristol Cyber Security group\\
	University of Bristol, UK\\
	\\	
 \And
	Awais Rashid \\
	Bristol Cyber Security group\\
	University of Bristol, UK\\
	\\		
}
\begin{document}


\title{Contrasting global approaches for identifying and managing cybersecurity risks in supply chains}

\begin{abstract}
Supply chains are increasingly targeted by threat actors. Using a recent taxonomy, we contrast the diverse levels of detail given by national authorities. The threat is commonly acknowledged, but guidance is disjointed. NIST SP 800-161 aligns closely with the taxonomy and offers a potential pathway towards a common set of principles. 

\end{abstract}
\maketitle

Cybersecurity Supply Chain Risk Management (C-SCRM) has gained strategic importance for organizations of all sizes and sectors globally. As organizations matured in their cybersecurity posture, threat actors needed new avenues to promote their nefarious activities. The evolution of digitization and inter-connectivity of the Information and Communications Technology (ICT) product and service supply chains has introduced a vector of attack that has seen rich pickings of late \cite{crowdstrike_crowdstrike_2022}.

Our previous research produced a C-SCRM taxonomy that is divided into four categories; Ownership, Risk, Service, and End-2-End.  \cite{topping_beware_2021}. The taxonomy allows for a rapid international comparison of the current threat landscape and the associated risk management guidance to understand whether there is commonality towards establishing C-SCRM processes, procedures, and practices. Such an approach is important as supply chains become increasingly global in nature and transcend national borders. It would also allow suppliers to comply with broadly common criteria, as opposed to a smorgasbord of national and sectorial conditions that may create additional complexities and potentially introduce conflicting requirements. 

The NotPetya cyber attack by Russia on Ukraine in 2017 and the more recent SolarWinds exploit \cite{lazarovitz_deconstructing_2021} were deployed through the compromise of legitimate software at source, with the client/supplier trust then used to deploy the malware.  

Incidents like these and the recent Apache Log4Shell/Log4j vulnerability \cite{zugec_technical_nodate} highlight the increased use of the supply chain in general, and the software supply chain in particular, as becoming a critical new frontier within cybersecurity. Organizations want to be able to understand the supply chain risk and to receive common guidance on how to mitigate it. Our review of North American, European, and Asian Pacific (APAC) national organizations found a large discrepancy in the level of detail provided at a national level, ranging from Highly Detailed to Minimal levels of advice and guidance for both the threat landscape and risk management.

For our comparison, we looked at major regions and focused on countries with highly developed cyber security national approaches. This included the USA and Canada for North America, whilst APAC was represented by Australia, New Zealand, and Singapore. For Europe, we included the UK, Germany, France, The Netherlands and, more widely, the European Union Agency for Cybersecurity (ENISA).

We also identified the US National Institute of Standards and Technology (NIST) as being a possible resource that authorities and organizations could gravitate towards for both detailed risk management  \href{https://doi.org/10.6028/NIST.SP.800-161r1}{NIST SP 800-161}  and a supporting  \href{https://nvlpubs.nist.gov/nistpubs/CSWP/NIST.CSWP.04162018.pdf}{cybersecurity framework}. This would support the goal of enabling a set of commonly defined principles that could then be adopted within specific national and sectoral operational contexts. 

\section{Supply Chain Threat Landscape}

\subsection{Regional and National Threat Assessments}
Accepting the trend of increased targeting of the cyber supply chain, we looked for progressive approaches taken to identify and communicate the threats and how they influence their guidance.

\begin{figure}
    \centering
    \includegraphics [width=0.50\textwidth]{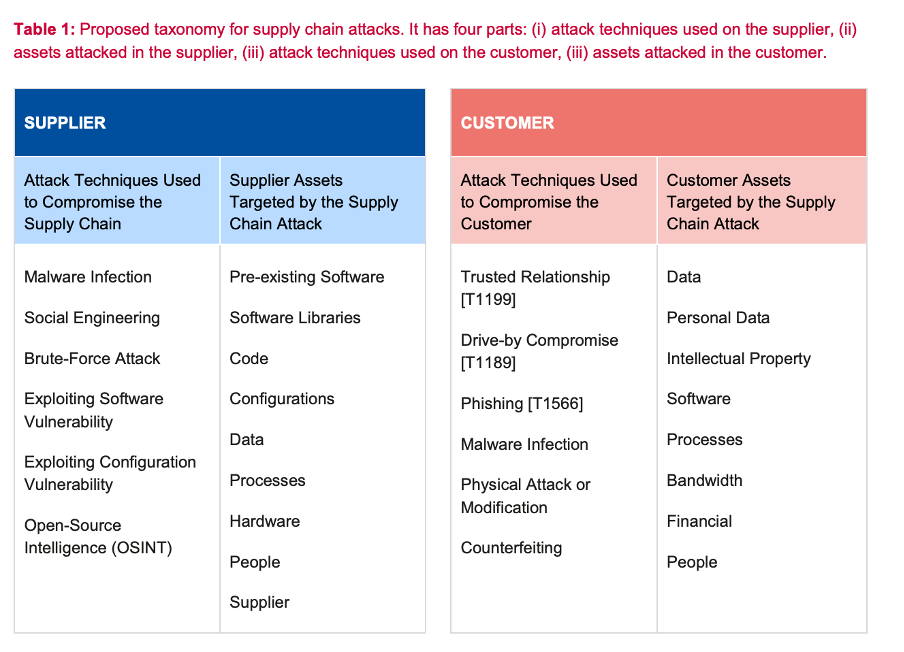}
    \caption{ENISA: Proposed taxonomy for supply chain attacks \cite{european_union_agency_for_cybersecurity_enisa_2021}}
    \label{fig:ENISA}
\end{figure}

The ENISA, in their recent report on the \href{https://data.europa.eu/doi/10.2824/168593}{Threat Landscape for Supply Chain Attacks}, mapped and studied supply chain attacks between January 2020 and July 2021. This period included the SolarWinds attack~\cite{lazarovitz_deconstructing_2021} and, more recently, the Kaseya ransomware attack~\cite{noauthor_kaseya_nodate}. The report develops a taxonomy (Figure \ref{fig:ENISA}) to classify supply chain attacks for improved analysis. Half of the 24 supply chain attacks studied were attributed to well-known Advanced Persistent Threat (APT) groups. 62 percent of attacks took advantage of the customers' trust in their suppliers; 66 percent focused on the suppliers' code as the vector of compromise, with data (customer, personal, and intellectual property) being the most popular target at 58 percent.

The UK NCSC predicted the increased vulnerability of the cyber supply chain to compromise in its \href{https://www.ncsc.gov.uk/news/annual-review-2018}{2018 Annual Review}. The report included reference to \textit{``the role the supply chain plays in leaving organizations vulnerable to compromise.''}. It evidenced supply chain compromises for both Managed Service Providers (MSPs) and legitimate software vendors. Compromise of the former would potentially give access to commercially sensitive data of the MSP's clients or access to their clients' networks via the trusted relationship, whilst targeting the latter enables a compromise of a product used by multiple organizations. 
Similar to many of the other organizations we reviewed, it maintains this focus on supply chain threats in subsequent annual reviews, whilst also providing both high-level and detailed advisories relating to critical supply chain vulnerabilities or compromises.

The APAC output is limited to contents within the respective annual cyber threat reports. Australia provides the \href{https://www.cyber.gov.au/sites/default/files/2021-09/ACSC%20Annual%20Cyber%20Threat%20Report%20-%202020-2021.pdf} {greatest details} of the cohort, focusing on the types of threat actors that target Australian organizations whilst also highlighting specific campaigns that were reliant upon supply chain attacks, such as SolarWinds. 
Expecting similar supply chain attacks to become prevalent, it focuses on software supply chain compromises and outlines the difficulty in detecting such attacks. It recommends organizations focus on incident response plans, almost with the assumption that compromise at some level is inevitable.

The US Cybersecurity and Infrastructure Security Agency (CISA) launched the ICT SCRM Task Force in October 2018. It was formed as a public-private partnership to provide advice and guidance on a number of working groups (WG). WG2 focuses on supplier, product, and service threat evaluation. The \href{https://www.cisa.gov/sites/default/files/publications/ict-scrm-task-force-threat-scenarios-report-v3.pdf}{latest version} included nine threat categories:

\begin{itemize}
    \item Counterfeit Parts
    \item External Attacks on Operations and Capabilities
    \item Internal Security Operations and Controls
    \item System Development Life Cycle (SDLC) Processes and tools
   \item Insider Threat
   \item Economic Risks
   \item Inherited Risk (Extended Supply Chain)
   \item Legal Risks
   \item External End-to-End Supply Chain Risks (Natural Disasters, Geo-Political Issues)
\end{itemize}

The report builds out a detailed threat list below these nine common threat groups and identifies associated threat categories and sources. It then creates threat scenarios against each threat group. This granular level of detail is ideal for a specialist team, but a higher-level document is required for the business senior stakeholders engaged in C-SCRM strategic planning.

There is a broad agreement amongst the organizations compared that the risks encompass third party products and services throughout the life of the contract. They routinely identify and provide tactical guidance on critical vulnerabilities that threat actors may target (such as Log4J) or active campaigns (such as SolarWinds). The level of detail differs for authorities, but the main crux is the appreciation of the threat that the cyber supply chain poses, and this should drive national authorities to promote best practice C-SCRM guidance that organizations large and small in all sectors can adopt.

\begin{figure*}[thtp]
    \centering
    \includegraphics[width=\textwidth]{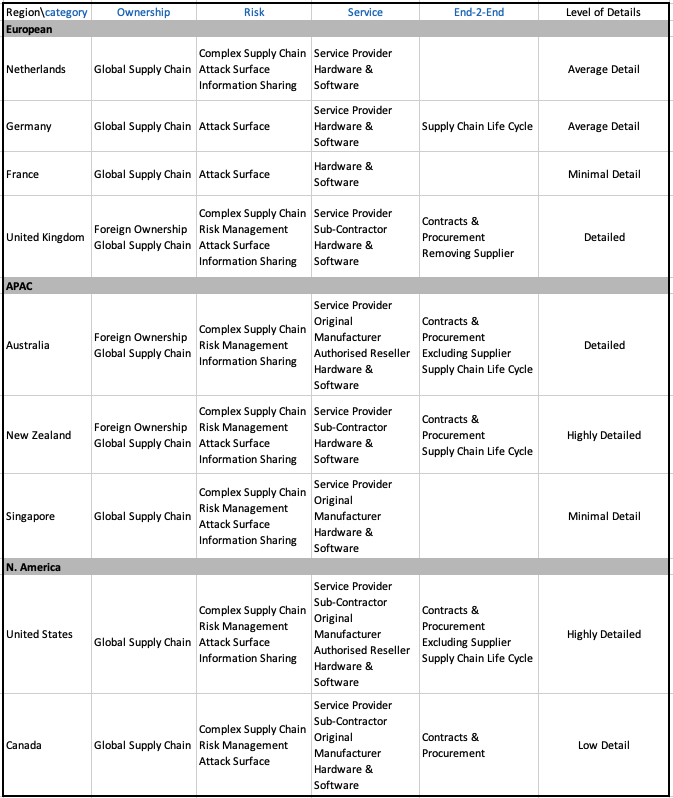}
    \caption{Regional and National C-SCRM Guidance}
    \label{fig:my_Review}
\end{figure*}

\section{Cybersecurity Supply Chain Risk Management}

The C-SCRM Taxonomy categories in Figure \ref{fig:my_Review} allowed us to compare the different regional and national levels of guidance to identify the sub-categories commonly addressed against those receiving less focus. Those sub-categories comprise the following:
\begin{itemize}
    \item \textbf{Ownership:} Foreign Ownership, Global Supply Chain, Private Entity.
    \item \textbf{Risk:} Complex Supply Chain, Risk Management, Attack Surface, Information Sharing.
    \item \textbf{Service:}Service Provider, Sub-Contractor, Original Manufacturer, Authorised Reseller, Hardware and Software.
    \item \textbf{End-2-End:} Contracts and Procurement, Excluding Supplier, Removing Supplier, Supply Chain Life Cycle.
\end{itemize}

We determined the levels of detail offered using the following five-point scale:

\begin{itemize}
    \item \textbf{Highly Detailed:} A clear level of detail across a broad spectrum of the sub-categories.
    \item \textbf{Detailed:} A broad spectrum was covered, but to a lesser degree of detail.
    \item \textbf{Average:} Less coverage, but to an adequate level of detail.
    \item \textbf{Low Detail:} Limited subset of categories/sub-categories covered, but with little detail.
    \item \textbf{Minimal:} Although a sub-category is referenced, there is no detail or definition.
    \item \textbf{None}. 
\end{itemize}

\subsubsection{Ownership}

All national authorities reviewed acknowledged that the supply chain is a global entity and that any compromise can transcend national boundaries. The UK is unique in the European cohort in addressing foreign ownership within the cyber supply chain and the associated risk that this may create. It extends that to also considering if data or services are offshored and the implications if that were the case. 

The Australia Cyber Security Centre (ACSC) also focuses heavily on foreign control, influence, and interference and notes that it can be difficult to identify the nationality of suppliers, particularly for multinational corporations. Similarly, the New Zealand NCSC points to regulatory and compliance requirements being clearly defined when employing international suppliers. It recommends a security review if a critical supplier has a change of ownership or major shareholding, or undergoes a merger. 

Although the Task Force WG4 has created a highly detailed vendor SCRM template, CISA doesn't cover foreign ownership or investment in the recommended controls. 

\subsubsection{Risk}
The Dutch NCSC's basic cybersecurity guidance doesn't reference supply chain risk management. This is despite identifying that small and medium-sized enterprises (SMEs) often fail to have adequate security controls, yet can form part of the supply chains of critical processes. It argues that a lack of transparency means that third-party risk assessments offer no guarantees of misuse. 
France provides tactical advice on critical vulnerabilities and compromises of hardware and software platforms, which is in line with all other national authorities assessed, but there is no evidence of C-SCRM advice. 

The UK provides good detail on securing the supply chain with a number of guidance documents. It covers similar ground to others around the attack surface that is created through a complex supply chain, allowing vulnerabilities to be inherent. This leads to risk management guidance offered at a high level, but starts with understanding one's supply chain and critical suppliers.

ACSC offers guidance on most of the risk management attributes, albeit with less detail than some. It calls for cybersecurity incident reporting to be open and transparent to customers and appropriate authorities in a timely manner. New Zealand noted that traditional IT is increasingly circumvented as business units deal directly with cloud and SaaS providers, thus introducing risk through the side door. 

CISA is aligned with most of those reviewed in acknowledging the complexity of the global digital supply chain and associated attack surface. There is a gap in guidance on information sharing, but increased detail regarding risk management advice, especially relating to business resilience strategies, and towards assurance and compliance. 

\subsubsection{Service}

Both the Dutch and UK NCSCs highlight outsourcing services such as cloud and SaaS as being a risk, although the Dutch do not currently offer any mitigation guidance. It does, however, note the misuse of Industrial Control System (ICS) supply chains and the increasing exploitation of vulnerabilities in hardware and software.

The German BSI focuses on the common practice of software projects, including software libraries that are reliant on others to maintain and that, if compromised, can impact countless software platforms. It is promoting the use of Software Bill of Materials (SBOMs) as a risk control measure. \href{http://www.bgbl.de/xaver/bgbl/start.xav?startbk=Bundesanzeiger_BGBl&jumpTo=bgbl121s1122.pdf} {The IT Security Act 2.0}  subjects original manufacturers of critical components (both hardware and software) to certain obligations that will flow down the supply chain to protect critical infrastructure.

APAC agencies cover similar ground. ACSC defines the supply chain as comprising the supplier, manufacturer, distributor, and retailer, as well as the service provider. It calls for a commitment to secure-by-design products using secure coding practices. Access control from vendors and service providers is also covered.

CISA provides a broad and detailed question set within this category, with most of the sub-categories and attributes covered multiple times. The requirement for products to be secure by design is particularly detailed, as is the life cycle maintenance and support.

\subsubsection{End-2-End}

Germany supports European certification schemes, such as consumer and public Internet of Things (IoT) devices and the need for an entire life cycle across its supply chain, whilst the UK calls for security controls and assurance to be contracted and to include the removal of the supplier, be that at contract end or through the termination of the contract. It was unique in the region to reference the exclusion of suppliers. It recently published \href{https://www.cpni.gov.uk/protected-procurement}{procurement guidance for business leaders, practitioners, and suppliers}. It mirrors a similar approach taken by New Zealand, which also signposts the need for contract cancellation provisions that ensure the return or disposal of assets and data from the supplier.

Cybersecurity expectations should be clearly documented within contracts and be justifiable, achievable, and proportionate to the risk exposed by the supplier, advises ACSC. Some suppliers may be excluded from receiving contracts if the government deems them a national security concern, particularly for critical infrastructure providers. Consideration for the supply chain life cycle is referenced repeatedly.

Singapore does not currently offer any specific C-SCRM guidance and didn't have sufficient output to populate the individual categories. However, it is advanced in implementing \href {https://www.csa.gov.sg/Programmes/certification-and-labelling-schemes/cybersecurity-labelling-scheme/about-cls} {security ratings for consumer IoT devices}.

Assurance that products are not procured from excluded suppliers is included in CISA guidance, although it doesn't address the removal of a supplier, be that at the natural end or termination of a contract. The supply chain lifecycle does have multiple control questions within the template.

The only C-SCRM guidance from the Canadian Centre for Cyber Security (CCCS) is limited and provides little detail.

\subsection{Threat Landscape and Risk Management comparison}

This rapid review reveals that one cannot assume that the level of detail an authority gives to the threat landscape will be similarly reflected in their risk management output. Most authorities are closely aligned in their product output, but there are exceptions. 

Whilst ENISA has produced the detailed \textit{Threat Landscape for Supply Chain Attacks}, we found no advice on how to manage that risk. It may be delegated down to individual member states, although, currently, the guidance coming from those authorities does not reflect this. The supply chain is covered within the  \href{https://eur-lex.europa.eu/legal-content/EN/TXT/HTML/?uri=CELEX:32016L1148&from=EN}{NIS Directive}. This would indeed then defer to member states, but only for operators of essential services and digital service providers. 

Conversely, New Zealand covers the threat landscape at a minimal level of detail, whilst their risk management guidance is highly detailed. 

The US provided the highest level of detail across the two areas. In general, there is slightly more detail in risk management than in threat awareness amongst national authorities. This would be expected as threat awareness offers general awareness, whilst risk management must provide specifics on how to address the threat.

All organizations reviewed offered some level of risk management advice, even if it was not particularly detailed. UK guidance comes from several documents. New Zealand and the US give very detailed advice, albeit with different approaches. New Zealand goes for a more principles-based assurance approach, whilst CISA outlines each possible control. This may be an issue if the template issued by CISA doesn't cover some of the taxonomy sub-categories.

\subsubsection{Coverage of the C-SCRM Taxonomy}
Nine national organizations were included in our review of their guidance against the C-SCRM taxonomy. Any sub-categories that were referenced seven or more times were deemed to be popular and are in bold font in Figure \ref{fig:my_Review}. The first thing to note is that there was no reference of the \textbf{Private Entity} sub-category within \textit{Ownership}. This attribute relates to shareholder and investor priorities. There are different business priorities between public and private organizations. This was acknowledged by the US and UK water sectors and by most of the frameworks in our previous paper \cite{topping_beware_2021}, but ignored at an authority level, and that has been replicated here.

All categories received at least one popular sub-category, with the exception of \textit{End-to-End}. Embedding C-SCRM into current and future contracts is key to transparency and accountability. This is still in its infancy for many organizations, although guidance and coverage are maturing. This will drive improvements in the supply chain lifecycle that will include the off-boarding of suppliers and service providers at contract end or termination.

Coverage of the \textbf{Global} and \textbf{Complex supply chain} are closely aligned and widely reported. This lends itself to the appreciation of the \textbf{Attack Surface} that threat actors target. This often involves the \textbf{Hardware, Software, and Service Providers}, so it is understandable that these areas receive good coverage.

Just because a national authority covers many of the sub-categories doesn't mean that they are providing quality advice. The final column provides the level of detail that we identified from the documents reviewed. It is the substance of the guidance that defines the level of detail provided.

\subsubsection{The search for clearly defined principles}

There needs to be synergy in C-SCRM to allow businesses to be exposed to the best and most appropriate authoritative advice and guidance, but also to allow supply chains and service providers to offer consistent assurances, regardless of the country or sector they support. 

National authorities are increasingly identifying NIST as the institution to support businesses in understanding and defining their cybersecurity posture, and this includes C-SCRM guidance. Although a US government agency, the guidance is freely available globally and transcends international borders. Their products also often map to complementary standards.

\subsection{NIST Standards}

\subsubsection{NIST SP 800-53 Rev 5 }
Titled "\textit{Security and Privacy Controls for Information Systems and Organizations}"; \href{https://doi.org/10.6028/NIST.SP.800-53Ar5}{Revision 5} was enhanced to include a number of controls and enhancements that relate to C-SCRM, including:
\begin{itemize}
    \item SCRM Control Family (SR)
    \item SCRM Strategy (PM-30)
    \item Suppliers of Critical or Mission Essential Items (PM-30(1))
    \item Integrated Situational Awareness (SI-4(17))
    \item External System Awareness
    \item Acquisition Process
    \item Supply Chain Risk Assessment
    \item Criticality Analysis
    \item Incident Handling - Supply Chain Coordination (IR-4(10))
    \item Incident Reporting - Supply Chain Coordination (IR-6(3))
    \item Controlled Maintenance (MA-2)
    \item Tamper Protection (PE-3(5))

\end{itemize}

It forms a key component of the Risk Management Framework (RMF) that also incorporates  \href{https://doi.org/10.6028/NIST.SP.800-37r2}{NIST SP 800-37}. 

\subsubsection{NIST SP 800-161 Rev 1}
Whilst NIST publications have elements of C-SCRM embedded within them, the newly published  \href{https://doi.org/10.6028/NIST.SP.800-161r1}{NIST SP 800-161 Rev 1} is the go-to document for "\textit{C-SCRM Practices for Systems and Organizations}". It integrates C-SCRM into risk management activities through a multi-level approach that includes the development of strategic implementation plans, C-SCRM policies, and risk assessments for products and services that apply to both IT and Operational Technology (OT) environments, and inclusive of IoT. It promotes the inclusion of a wide array of stakeholder groups and covers activities that span the entire System Development Life Cycle (SDLC). 

It references NIST publications and legislative developments that have influenced the content. It also references international standards, guidelines, and best practice documents, the majority of which are provided by the International Organization for Standardization and the International Electrotechnical Commission (ISO/IEC). 

When compared against the taxonomy, NIST SP 800-161 is broadly aligned, even down to the attribute level. Some gaps may be covered by other NIST documents. \textit{Exclusion of suppliers} is not directly considered, but it does indicate adherence to legislative requirements, where this would be covered. The \textit{removal of a supplier} is only briefly covered. These are key components of C-SCRM and would benefit from being more obviously signposted. 

There is no consideration that service providers have multiple customers, which may include direct competitors and require contractual separation of duties. Additionally, the business priorities of public and private organizations are often very different. Promoting that awareness would afford a greater appreciation of that potential dichotomy.

Despite these limitations, coupled with the observation that NIST is a US government agency focused on promoting US innovation and industrial competitiveness, the content is highly detailed and inclusive for an international audience. This lends itself to becoming the pathway towards a common set of principles. It aligns closely with the C-SCRM taxonomy, addressing most of the gaps identified in this rapid review of global guidance. This allows for a complementary approach whereby the taxonomy promotes an appreciation of what is required, whilst NIST can provide the systemic approach to addressing those needs. This will allow regional authorities to deliver similar guidance at a level of detail that satisfies both their mission statement and their audience needs.

\section{A glimmer of hope!}

The challenge remains for national authorities to come together to speak with one authoritative voice. On May 11, 2022, Australian, British, Canadian, New Zealand, and United States agencies jointly issued a \href{https://www.cisa.gov/uscert/ncas/alerts/aa22-131a}{cybersecurity advisory} to protect MSPs and customers. Although restricted to traditional political "5-Eyes" partners, it hopefully announces the start of more collaborative approaches towards a global common set of C-SCRM principles.


\printbibliography[title=References]

@online{crowdstrike_crowdstrike_2022,
	title = {Crowdstrike 2022 Global Threat Report},
	url = {https://go.crowdstrike.com/rs/281-OBQ-266/images/Report2022GTR.pdf},
	author = {{Crowdstrike}},
	urldate = {2022-04-02},
	date = {2022-02},
}

@online{zugec_technical_nodate,
	title = {Technical Advisory: Zero-day critical vulnerability in Log4j2 exploited in the wild},
	url = {https://businessinsights.bitdefender.com/technical-advisory-zero-day-critical-vulnerability-in-log4j2-exploited-in-the-wild},
	shorttitle = {Technical Advisory},
	author = {Zugec, Martin},
	urldate = {2021-12-18},
	langid = {english},
}

@article{topping_beware_2021,
	title = {Beware suppliers bearing gifts!: Analysing coverage of supply chain cyber security in critical national infrastructure sectorial and cross-sectorial frameworks},
	volume = {108},
	issn = {01674048},
	url = {https://linkinghub.elsevier.com/retrieve/pii/S0167404821001486},
	doi = {10.1016/j.cose.2021.102324},
	shorttitle = {Beware suppliers bearing gifts!},
	pages = {102324},
	journaltitle = {Computers \& Security},
	shortjournal = {Computers \& Security},
	author = {Topping, Colin and Dwyer, Andrew and Michalec, Ola and Craggs, Barnaby and Rashid, Awais},
	urldate = {2021-11-21},
	date = {2021-09},
	langid = {english},
}

@article{lazarovitz_deconstructing_2021,
	title = {Deconstructing the {SolarWinds} breach},
	volume = {2021},
	issn = {1361-3723},
	url = {https://www.sciencedirect.com/science/article/pii/S1361372321000658},
	doi = {10.1016/S1361-3723(21)00065-8},
	pages = {17--19},
	number = {6},
	journaltitle = {Computer Fraud \& Security},
	shortjournal = {Computer Fraud \& Security},
	author = {Lazarovitz, Lavi},
	urldate = {2021-11-20},
	date = {2021-06-01},
	langid = {english},
}

@online{noauthor_kaseya_nodate,
	title = {Kaseya Ransomware Attack: Guidance for Affected {MSPs} and their Customers {\textbar} {CISA}},
	url = {https://us-cert.cisa.gov/kaseya-ransomware-attack},
	urldate = {2021-11-20},
}

@book{european_union_agency_for_cybersecurity_enisa_2021,
	location = {{LU}},
	title = {{ENISA} threat landscape for supply chain attacks.},
	url = {https://data.europa.eu/doi/10.2824/168593},
	publisher = {Publications Office},
	author = {{European Union Agency for Cybersecurity.}},
	urldate = {2021-08-17},
	date = {2021},
}

\paragraph{Colin Topping} is the cyber incident director at Rolls-Royce plc. He is also undertaking a part-time Ph.D. at the Bristol Cyber Security Research Group, University of Bristol, United Kingdom. This is funded by the National Cyber Security Centre. His principal research interest is focused on cyber security within the supply chain in an ever increasing global, technical, and interdependent environment.

\paragraph{Ola Michalec} is a Senior Research Associate at the Bristol Cyber Security Research Group, University of Bristol, United Kingdom. Her interests span the design and implementation of infrastructure policies regarding climate mitigation and cyber security.

\paragraph{Awais Rashid} is professor of cyber security with the University of Bristol, United Kingdom. His research interests are in cyber security risk assessment, security of cyber-physical systems, and human and organizational aspects of security. He is Director of the EPSRC Centre for Doctoral Training in Trust, Identity, Privacy and Security in Large-scale Infrastructures, is a fellow of the Alan Turing Institute and leads projects as part of two UK Research Institutes: the Research Institute in Trustworthy Industrial Control Systems (RITICS), the Research Institute on Science of Cyber Security (RISCS) and the National Centre of Excellence on Cyber Security of IoT (PETRAS). He also heads a major international effort to develop a Cyber Security Body of Knowledge (CyBOK).

\end{document}